\documentclass[12pt]{article}

\def\slash#1{\setbox0=\hbox{$#1$}#1\hskip-\wd0\hbox to\wd0{\hss\sl/\/\hss}}
\usepackage{epsf, cite}
\usepackage{epsfig}
\usepackage[all]{xy}
\usepackage{amsfonts}
\usepackage{latexsym}

\usepackage{epsf, cite}

\makeatletter
\renewcommand\section{\@startsection {section}{1}{\z@}%
                                   {-3.5ex \@plus -1ex \@minus -.2ex}
                                   {2.3ex \@plus.2ex}%
                                   {\normalfont\large\bfseries}}
\renewcommand\subsection{\@startsection{subsection}{2}{\z@}%
                                     {-3.25ex\@plus -1ex \@minus -.2ex}%
                                     {1.5ex \@plus .2ex}%
                                     {\normalfont\bfseries}}
\makeatother

\let\non\nonumber

\def\lbldef#1#2{\expandafter\gdef\csname #1\endcsname {#2}}

\def\href#1#2{#2}

\newcommand{\beq}{\begin{equation}}
\newcommand{\eeq}{\end{equation}}

\def\bea{\begin{eqnarray}}
\def\eea{\end{eqnarray}}

\providecommand{\openone}{\leavevmode\hbox{\small1\kern-3.8pt\normalsize1}}
\def\non{\nonumber}
\def\xt{\tilde{X}}
\def\mt{\tilde{m}}
\def\wt{\tilde{w}}
\def\ut{\tilde{u}}
\def\vt{\tilde{v}}
\def\Lt{\tilde{L}}
\def\nt{\tilde{n}}
\def\ti{\tilde{I}}
\def\tj{\tilde{J}}
\def\ct{\tilde{c}}

\def\zb{\bar{z}}
\def\tb{\bar{\tau}}
\def\Z{{\mathbb Z}}
\def\mod{\ \mbox{mod}\ }

\def\bx{\square}

\def\P{\mathcal P}
\def\D{\mathcal D}
\def\H{\mathcal H}
\def\V{\mathcal V}
\def\X{\mathbb X}

\begin{document}

\setcounter{page}{1}



\begin{titlepage}

\begin{center}

\hfill QMUL-PH-06-14\\
\hfill DAMTP-2007-1\\

\vskip 2 cm
{\Large \bf
The string partition function in Hull's doubled formalism}\\

\vskip 1.25 cm {David S. Berman\footnote{email: D.S.Berman@qmul.ac.uk}}
\\
{\vskip 0.5cm
Queen Mary College, University of London,\\
Department of Physics,\\
Mile End Road,\\
London, E1 4NS, England.}\\
{\vskip 0.5cm
and}

\vskip 0.5 cm {Neil B. Copland\footnote{email: N.B.Copland@damtp.cam.ac.uk}}
\\
{\vskip 0.5cm
Department of Applied Mathematics and Theoretical Physics,\\
Centre for Mathematical Sciences,\\University of Cambridge,\\
Wilberforce Road,\\
Cambridge CB3 0WA,\\ England.}\\

\end{center}

\vskip 1 cm

\begin{abstract}
\baselineskip=18pt
T-duality is one of the essential elements of string theory. Recently,
Hull has developed a formalism where the dimension of the target space
is doubled so as to make T-duality manifest. This is then supplemented 
with
a constraint equation that allows the connection to the usual string
sigma model. This paper analyses the partition function of the doubled
formalism by interpreting the constraint equation as that of a chiral
scalar and then using holomorphic factorisation
techniques to determine the partition function. We find there is
quantum equivalence to the ordinary string once the topological
interaction term is included.

\end{abstract}

\end{titlepage}

\pagestyle{plain}

\baselineskip=19pt
\section{Introduction}

One of the crucial differences between string theory and other
theories of quantum gravity is T-duality. The quantum equivalence
between different target spaces in which the string propagates is a
fascinating consequence of string dynamics. Recently, Hull has
developed a formalism where this symmetry is manifest by doubling the
dimension of the target space so as to include all duality related
geometries in a single target space. The theory is then supplemented
with constraints so as to allow a reduction (through gauging) of 
the theory back to usual string string theory. The formalism has 
the advantage of providing a geometric manifestation of T-duality 
but more importantly perhaps it may be crucial in formulating string 
theory on non-trivial T-folds. The formalism we will use was developed in
\cite{Hull:2004in} and \cite{Hull:2006va} though based on earlier work 
\cite{tseytlin}. We should mention there have been a variety of papers 
which have studied T-folds \cite{tfold1} as possible string vacua; it is 
hoped the doubled formalism will be useful to issues arising in T-fold 
compactifications even though this paper will study the formalism on a 
non-fibred circle.

The goal of this paper is to study the quantum theory of the doubled
formalism. This has already been examined from the point of view of
canonical quantisation with a detailed constraint analysis by
Hacket-Jones and Moutsopoulos\cite{emily} also using results of Hellerman and Walcher\cite{walcher}.  

In this paper we will take the following viewpoint. The constraints are to be viewed as a those determining 
chiral scalars. The goal is to show that
these chiral Bosons combine in the right way for us to interpret the partition
function for one chiral Boson and its dual as that of an ordinary non-chiral Boson. (These chiral Bosons
cannot be interpreted quite in the usual way since they have novel
periodicity conditions, as will be seen below).

The quantum theory of chiral Bosons is an extraordinarily subtle
business. A host of publications contain the problems and methods of
dealing with them \cite{Witten} but perhaps a special mention should
go to \cite{moore} where the details of how to calculate the partition
function are explained in detail. We will follow many of the 
techniques explained
there (see the Appendices of \cite{moore}). The main subtlety arises
in the treatment of the sum over instantons, i.e. the sum over the 
cohomological part of the Bosons.

The method is as
follows. To calculate the partition function of a chiral scalar one
must first calculate the partition function of the non-chiral scalar and
then carry out a variety of resummations and manipulations so as to
write it as a product of a holomorphic
function and an anti-holomorphic function. One then identifies the
holomorphic part as the chiral scalar partition function. Essentially,
one is factorising the usual partition function into contributions
from the chiral and anti-chiral pieces. (We have skirted over
the details of picking spin structures and importantly for our
calculation the factorisation is more problematic when the Boson is 
not at the free-Fermion radius.) 

In what follows we calculate the
partition function of scalars on the doubled torus. Then we carry out a
number of non-trivial resummations so as to factorise the partition
function into pieces that may be interpreted as the contribution from
the chiral modes obeying Hull's constraint. 
These pieces are then recombined to give the usual string
partition function.

\section{Factorising the Partition Function}

We begin with the doubled string action for a single toroidally
compactified Boson, so that we have the Boson $X$, 
which is associated to a circle of radius $R$, and its dual $\xt$ which is associated to a circle of radius $R^{-1}$. The action is given by
\beq\label{Xact}
S=\frac{\pi}{2} R^2dX\wedge* dX+\frac{\pi}{2} R^{-2}d\xt\wedge* d\xt.
\eeq
The constraint in this geometry (see Appendix) can be written in a simple form in terms of the 
linear combinations
\beq\label{PQ}
P=RX+R^{-1}\xt\, ,\qquad Q=RX-R^{-1}\xt.
\eeq
Note that the normalisation is a factor of $1/2$ times the usual one. 
This was the normalisation needed by Hull in \cite{Hull:2006va} when 
gauging the current associated with the constraint to show equivalence
with the standard sigma model. Also, here $R=1$ corresponds to the T-duality self-dual radius, whereas in much of the literature on holomorphic factorisation $R=1$ is the free-Fermion radius (which would be $R=1/\sqrt{2}$ in our conventions). A topological term given by 
\beq
L_{top}=\pi dX\wedge d\xt\, 
\eeq
was also required so as to have invariance under large gauge
transformation when gauging the current. We also find the need to
include it here and in fact it is crucial in what follows for 
demonstrating quantum equivalence.

The constraints are 
\beq
\partial_{\zb} P=0\, , \qquad\partial_{z} Q=0,
\eeq
that is $P$ is holomorphic and $Q$ is anti-holomorphic. 
We can rewrite the action in terms of $P$ and $Q$, as follows
\beq
S=\frac{\pi}{4} dP\wedge* dP+\frac{\pi}{4} dQ\wedge* dQ.
\eeq
The standard way of obtaining a partition function for a chiral Boson
is to factorise the partition function for an ordinary Boson into
holomorphic and anti-holomorphic parts and keep only the factor with
the correct holomorphic dependence \cite{moore}. Here we would like to do
this for $P$ and $Q$. However, crucially $P$ and $Q$ do not have standard
periodicity properties, and are linked through (\ref{PQ}). This is why
we cannot directly identify these fields as left and right movers on a
circle; they do not have the right periodicity conditions.

We will now examine the {\it{instanton}} sector. This means we need
to examine contributions to the partition function which, from the field 
theory point of view, will be in the cohomological sector.
When writing the action (\ref{Xact}), $dX$ should really be replaced by $L$, where $L=dX+\omega$, with $\omega\in
H^1(\Sigma,\Z)$. 
We can express the cohomological part in terms of the standard cohomology basis, which
for a toroidal worldsheet consists of just one $\alpha$ cycle and one $\beta$ cycle. 
\beq
L=dX+n\alpha+m\beta\, ,\qquad \Lt=d\xt+\nt\alpha+\mt\beta,
\eeq
where $n,\nt,m,\mt\in \Z$, which in turn means that we should replace $dP$ and $dQ$ by
\beq
M=dP+(Rn+R^{-1}\nt)\alpha+(Rm+R^{-1}\mt)\beta\, ,\qquad N=dQ+(Rn-R^{-1}\nt)\alpha+(Rm-R^{-1}\mt)\beta\,.
\eeq
The classical or {\it{instanton}} sector of the partition function is
the sum over all field 
configurations of $\exp[-S]$ and can be written as
\bea\label{Rpf}
Z&=&\sum_{m,n,\mt,\nt} \exp \left[ -(Rm+R^{-1}\nt)^2\frac{
\pi|\tau|^2}{4\tau_2}+(Rn+R^{-1}\nt)(Rm+R^{-1}\mt)\frac{\pi\tau_1}{2\tau_2}-(Rm+R^{-1}\mt)^2\frac{\pi}{4\tau_2}\right]\nonumber \\ 
&&\times \exp \left[ -(Rm-R^{-1}\nt)^2\frac{\pi|\tau|^2}{4\tau_2}+(Rn-R^{-1}\nt)(Rm-R^{-1}\mt)\frac{\pi\tau_1}{2\tau_2}-(Rm-R^{-1}\mt)^2\frac{\pi}{4\tau_2}\right]\, ,
\eea
where the first factor corresponds to $P$ and the second to $Q$. 
To holomorphically factorise the partition function we must Poisson
resum, but first we separate the sums so that the $P$ 
and the $Q$ parts of the partition function sum over different 
independent variables. The contribution from 
the topological term within the sum is
\beq
\exp[i\pi (n\mt -m\nt)].
\eeq
This will only contribute a sign change to the terms in the partition
sum, however, this is crucial in showing the equivalence to the standard formulation.

To be able to separate the sums we assume $R^2=\frac{p}{q}$, with $p,q$ coprime integers, and let $k=pq$. Then we have
\beq
Rn\pm R^{-1}\nt=\sqrt{k}\left(\frac{n}{q}\pm\frac{\nt}{p}\right).
\eeq
Making the substitutions $n=cq+q\gamma_q$ and $\nt=\ct p+p\gamma_p$ (where $c,\ct\in\Z$ and $\gamma_q\in \{0,\frac{1}{q},\ldots,\frac{q-1}{q}\}$) we can further say
\beq
\sqrt{k}\left(\frac{n}{q}\pm\frac{\nt}{p}\right)=\sqrt{k}(c\pm\ct +\gamma_q\pm \gamma_p).
\eeq
Then we let $h=c+\ct$ and $l=c-\ct$. We have rewritten the sum over
$n$ and $\nt$ as a sum over $c,\ct,\gamma_q$ and $\gamma_p$, and then 
rewritten the $c$ and $\ct$ sums as a sum over $h$ and $l\in\Z$, 
but $h-l=2\ct$ so we must restrict to even values of $h-l$ by inserting a factor of 
\beq
\sum_{\phi\in\{0,\frac{1}{2}\}}\frac{1}{2}\exp[2\pi i \phi(h-l)]
\eeq
in the partition function. Repeating the process to split the $m,\mt$ sum, the partition function becomes
\bea\label{pqpf}
Z=\sum_{\phi,\theta,\gamma_q,\gamma_p,\gamma'_q,\gamma'_p}&& \sum_{h,
i,j,l}\frac{1}{4}\exp\left[-\frac{k\pi}{4}\left\{(h+\gamma_q+\gamma_p)^2\frac{|\tau|^2}{\tau_2}-2(h+\gamma_q+\gamma_p)(i+\gamma'_q+\gamma'_p)\frac{\tau_1}{\tau_2}\right.\right.\nonumber \\ 
&&\left.+(i+\gamma'_q+\gamma'_p)^2\frac{1}{\tau_2}\right\}-\frac{k\pi}{4}\left\{(l+\gamma_q-\gamma_p)^2\frac{|\tau|^2}{\tau_2}\right.\nonumber\\
&&\left.-2(l+\gamma_q-\gamma_p)(j+\gamma'_q-\gamma'_p)\frac{\tau_1}{\tau_2}+(j+\gamma'_q-\gamma'_p)^2\frac{1}{\tau_2}\right\}\nonumber \\ 
&&+2\pi i\left\{\phi(h-l)+\theta(i-j)\right\}\non\\
&&+\left.\frac{i\pi k}{2} \left((l+\gamma_q-\gamma_p)(i+\gamma'_q+\gamma'_p)-(h+\gamma_q+\gamma_p)(j+\gamma'_q-\gamma'_p)\right)\right].
\eea
Using the notation $\gamma_\pm=\gamma_q\pm\gamma_p$ we can rewrite the partition function again as
\bea\label{PQpf}
Z=\sum_{\phi,\theta,\gamma_q,\gamma_p,\gamma'_q,\gamma'_p}\sum_{h,
i,j,l} \left(\frac{1}{2}\exp \right.&&\left[-\frac{k\pi}{4}\left\{(h+\gamma_+)^2\frac{|\tau|^2}{\tau_2}-2(h+\gamma_+)(i+\gamma'_+)\frac{\tau_1}{\tau_2}+(i+\gamma'_+)^2\frac{1}{\tau_2}\right\}\right. \nonumber\\
&&+2\pi i\left\{\phi h+\theta i\right\}\Biggr]\nonumber \\ 
\times\frac{1}{2}\exp&&\left[-\frac{k\pi}{4}\left\{(l+\gamma_-)^2\frac{|\tau|^2}{\tau_2}-2(l+\gamma_-)(j+\gamma'_-)\frac{\tau_1}{\tau_2}+(j+\gamma'_-)^2\frac{1}{\tau_2}\right\}\right. \nonumber\\
&&-2\pi i\left\{\phi l+\theta j\right\}\Biggr]\nonumber\\
\times\exp&&\left.\left[\frac{i\pi k}{2} \left((l+\gamma_-)(i+\gamma'_+)-(h+\gamma_+)(j+\gamma'_-)\right)\right]\right)\, ,
\eea
where we have split the terms in the partition sum into three factors, the piece coming from the $P$
kinetic term (which depends on $h$ and $i$), the piece coming from the
$Q$ kinetic term (which depends on $l$ and $j$), and the topological
piece, a cross term depending on all four integer indices.

We now wish to perform Poisson resummation on $i$ and $j$. 
Let us focus on the $i$ resummation, replacing it with a sum over
$r$. Here we rewrite the $P$ part of the partition function and the first term of the topological piece:
\bea\label{Ppf}
Z_P&=&\sum_{\phi,\theta,\gamma_q,\gamma_p,\gamma'_q,\gamma'_p}\sum_{h,l,
r} \frac{1}{2}\sqrt{\frac{4\tau_2}{k}}\exp \Biggl[ -\frac{k\pi}{4}\left\{(h+\gamma_+)^2\frac{|\tau|^2}{\tau_2}-2\gamma'_+(h+\gamma_+)\frac{\tau_1}{\tau_2}+(\gamma'_+)^2\frac{1}{\tau_2}\right\} \nonumber\\
&&\qquad\qquad\qquad+2\pi i\phi h+\frac{i\pi k}{2}(l+\gamma_-)\gamma'_+\non\\&&
\qquad\qquad\qquad-\frac{4\pi\tau_2}{k}\left(r-\theta+ik\frac{(h+\gamma_+)}{4}\frac{\tau_1}{\tau_2}-\frac{ik\gamma'_+}{4\tau_2}-\frac{k}{4}(l+\gamma_-)\right)^2\Biggr]\\
&=&\sum_{\phi,\theta,\gamma_q,\gamma_p,\gamma'_q,\gamma'_p}\sum_{h,l,r}\sqrt{\frac{\tau_2}{k}}\exp \Biggl[ \tau_2\Biggl(-\frac{k\pi}{4}(h+\gamma_+)^2-4\pi k\left(\frac{r-\theta}{k}-\frac{1}{4}(l+\gamma_-)\right)^2\Biggr)\nonumber\\
&&-2\pi i \tau_1\left(h+\gamma_+\right)(r-\theta-\frac{k}{4}(l+\gamma_-))+2\pi i\phi h+2\pi i \left(r-\theta\right)\gamma'_+\Biggr].
\eea

The appearance of the squares with $\tau_2$ and the cross term with
$\tau_1$ is standard and Poisson resumming to replace $j$ with $s$ in
the $Q$ part of the partition function we can rewrite the whole partition function as
\bea
Z=\sum_{\phi,\theta,\gamma_q,\gamma_p,\gamma'_q,\gamma'_p,h,l,r,s}&\Bigl(\sqrt{\frac{\tau_2}{2k}}\exp \left[i\pi k\tau \frac{p_L^2}{2}-i\pi k\tb \frac{p_R^2}{2}+2\pi i  \left(\phi h+(r-\theta)\gamma'_+\right)\right]\nonumber \\ 
&\times\ \sqrt{\frac{\tau_2}{2k}}\exp \left[i\pi k\tau \frac{q_L^2}{2}-i\pi k\tb \frac{q_R^2}{2}+2\pi i  \left(-\phi l+(s+\theta)\gamma'_-\right)\right]\Bigr)
\eea
where
\bea\label{pqlr}
p_L&=&\frac{1}{2}\left(h+\gamma_+\right)-2\left(\frac{r-\theta}{k}-\frac{1}{4}(l+\gamma_-)\right)\, ,\non\\
p_R &=&\frac{1}{2}\left(h+\gamma_+\right)+2\left(\frac{r-\theta}{k}-\frac{1}{4}(l+\gamma_-)\right)\, , \non\\
q_L&=&\frac{1}{2}\left(l+\gamma_-\right)-2\left(\frac{s+\theta}{k}+\frac{1}{4}(h+\gamma_+)\right)\, ,\non\\
q_R &=&\frac{1}{2}\left(l+\gamma_-\right)+2\left(\frac{s+\theta}{k}+\frac{1}{4}(h+\gamma_+)\right)\, .
\eea
We can now see the clear holomorphic and anti-holomorphic
parts of the partition function for both $P$ and $Q$, as well as
additional pieces which restrict the sum over ``momenta'',
$p_L,p_R,q_L,q_R$. 
However, because the sums are linked we cannot remove the extra
pieces. So we rewrite the sums again, reconstructing them back to a sum over just four integers.

The $h,l$ terms in the momenta can easily be recombined just by undoing the substitutions we made earlier to write 
\beq 
h+\gamma_+=\frac{n}{q}+\frac{\nt}{p}\, , \qquad l+\gamma_-=\frac{n}{q}-\frac{\nt}{p}\, ,
\eeq
where we have replaced the sums over $h,l,\gamma_+,\gamma_-$ and
$\phi$ with a sum over integers $n$ and $\nt$, we also remove 
one of the factors of $\frac{1}{2}$ we inserted outside the partition function.

The other sum has been Poisson resummed so the recombination is more complicated. We use the identity
\beq\label{rootsof1}
\sum_{k=0}^{n-1}\left(\exp \left(\frac{2\pi i k}{n}\right)\right)^j=\sum_{\gamma_n}\exp (2\pi i \gamma_n j)=\left\{\begin{array}{l}
n\ \mbox{if}\ j \equiv 0 \mod n\\ 0\ \mbox{otherwise}
\end{array}\right. .
\eeq
With that in mind we see the only occurrence of $\gamma'_q$ and $\gamma'_p$ is in the factor
\bea
&&\sum_{\gamma'_q,\gamma'_p}\exp \left[2\pi i\left(\frac{(p_L-p_R)}{2}\gamma'_++\frac{(q_L-q_R)}{2}\gamma'_-\right)\right]\non\\
&=&\sum_{\gamma'_q,\gamma'_p}\exp\left[2\pi i (r+s)\gamma'_q+2\pi i(r-s-2\theta)\gamma'_p\right]
\eea
which has the effect of enforcing
\bea
r+s&\equiv& 0 \mod q\, ,\nonumber\\
r-s-2\theta&\equiv& 0 \mod p\, ,
\eea
in the rest of the partition sum. We see that these requirements are fulfilled exactly by putting
\beq
\frac{r-\theta}{k}=\frac{1}{2}\left(\frac{w}{p}+\frac{\wt}{q}\right)\, , \qquad \frac{s+\theta}{k}=\frac{1}{2}\left(\frac{w}{p}-\frac{\wt}{q}\right)\, .
\eeq
We have replaced the sums over $r,s,\theta,\gamma'_p$ and $\gamma'_q$
by a sum over integers $w,\wt \in\Z$. 
Note that importantly it is the term with $q$ in the denominator which
changes sign between the two combinations. 
Also, due to (\ref{rootsof1}) we get a factor of $k=pq$ outside the
exponential, which cancels the factor of $1/k$ obtained from Poisson resummation. 
We can now rewrite (\ref{pqlr}) as
\bea\label{pqlr2}
p_L=\frac{n}{q}-\left(\frac{w}{p}+\frac{\wt}{q}\right)\, ,\qquad p_R &=&\frac{\nt}{p}+\left(\frac{w}{p}+\frac{\wt}{q}\right)\, , \nonumber\\
q_L=-\frac{\nt}{p}-\left(\frac{w}{p}-\frac{\wt}{q}\right)\, ,\qquad q_R &=&\frac{n}{q}+\left(\frac{w}{p}-\frac{\wt}{q}\right)\, .
\eea

The doubled partition function is now in the simple form
\beq\label{Zd}
Z_d=\sum_{p_L,
p_R} \sqrt{2\tau_2}\exp\left[i\pi k\tau \frac{p_L^2}{4}-i\pi k\tb \frac{p_{R}^{2}}{4}\right] \sum_{q_L,q_R} \sqrt{2\tau_2}\exp\left[i\pi k\tau \frac{q_L^2}{4}-i\pi k\tb \frac{q_R^2}{4}\right].
\eeq

Now we can make the following final substitution
\bea
u=n-\wt\, ,\qquad v&=&-w\, ,\nonumber\\
\ut=\wt\, ,\qquad \vt&=&\nt+w\, ,
\eea
leading to  
\bea\label{pqlr3}
p_L=\frac{u}{q}+\frac{v}{p}\, ,\qquad p_R &=&\frac{\ut}{q}+\frac{\vt}{p}\, , \nonumber\\
q_L=\frac{\ut}{q}-\frac{\vt}{p}\, ,\qquad q_R &=&\frac{u}{q}-\frac{v}{p}\, .
\eea
It is the shift in the momenta caused by the topological term that
allows us to rewrite $n$ and $\wt$ 
in terms of {\it independent} summation variables $u , \ut ,v $ and $\vt$. 

The pieces of the appropriate holomorphicity that we wish to
keep are now summed over the same indices, and are not linked to the pieces
which we wish to remove. We may therefore remove the anti-holomorphic part 
of the partition function coming from $P$ (the $p_R$ piece) and the 
holomorphic part of the partition function coming from $Q$ (the $q_L$
piece). This leaves us with
\beq\label{Zf}
Z_f=\sum_{p_L,
p_R} \sqrt{2\tau_2}\exp\left[i\pi \tau \frac{p_L^2}{4}-i\pi\tb \frac{q_{R}^{2}}{4}\right]
\eeq
where
\beq
p_L=uR+\frac{v}{R}\, , \qquad q_R=uR-\frac{v}{R}\, .
\eeq
Alternatively we can see that the final form of the instanton part of the partition function, $Z_f$, is the holomorphic square root of the doubled contribution to the partition function:
\beq
Z_d=Z_f\times \bar{Z_f}.
\eeq
$Z_f$ is the instanton part of the partition function for a standard Boson with radius $R$
(or $R^{-1}$ by relabelling the indices), up to a factor outside the exponential. We will find that when we consider the rest of the partition function in the doubled formalism in the next section, the inverse of this factor will occur, giving an identical total partition function.

The approach taken here has been to treat the Bosons in the doubled formalism 
as chiral Bosons when trying to quantise. There is a key difference however, 
for chiral Bosons one must pick a spin structure \cite{Witten} and we have not 
done so here. If one were to do so then there would not be enough degrees of 
freedom to reconstruct the usual non-chiral Boson. 
Thus when one is holomorphically factorising here, one is effectively 
keeping a sum of chiral Bosons with all spin structures. This prescription is an essential part of the quantum prescription of the theory.

\section{The Oscillators}

So far we have only included the sum over solutions to the classical equations of motion, to complete the quantum path integral we must include the fluctuations around these classical solutions\cite{polchinski}. For a Boson $X$ with action
\beq
S=-\frac{\pi R^2}{2} dX\wedge *dX\, ,
\eeq
we must perform the Gaussian integral
\beq
\int\D X e^{-\int\frac{\pi R^2}{2}X \bx X},
\eeq
where $\bx$ is the Laplacian. The $\D X$ integration is split into the zero-mode piece and the integral over $\D X'$, orthogonal to the zero-mode. As $X$ has period 1 in our conventions, the zero-mode contribution is only a factor of 1. To normalise the measure we insert a factor of
\bea\label{norm}
&&\left(\int dx\, e^{-\frac{\pi R^2}{2}\int x^2}\right)^{-1}\non\\
&=&\left(\frac{\pi}{\frac{\pi R^2}{2}\int 1}\right)^{-1/2}\non\\
&=&\frac{R}{\sqrt{2}},
\eea
where we have used the fact that with our conventions $\int 1 $ over the torus is 1. 
This means
\beq
Z_{osc}=\frac{R}{\sqrt{2}}\frac{1}{\det' \bx}.
\eeq
We evaluate the determinant of $\bx=-4\tau_2\partial\bar{\partial}$ as a regularised product of eigenvalues, where the $'$ indicates this does not include zero-modes. We use a basis of eigenfunctions
\beq\label{efns}
\psi_{nm}=\exp \left[\frac{2\pi i}{2i\tau_2}\left(n(z-\zb)+m(\tau \zb-\tb z)\right)\right],
\eeq
which is single valued under $z\rightarrow z+1$ and $z\rightarrow z+\tau$, where $z=\sigma_1+\tau\sigma_0$ and $\tau$ is the complex structure of the toroidal worldsheet. The regularised determinant is then the product of eigenvalues
\beq
\mbox{det} ' \bx=\prod_{\{m,n\}\neq\{0,0\}}\frac{4\pi^2}{\tau_2}(n-\tau m)(n-\tb m)\,.
\eeq
This can be evaluated using $\zeta$-function regularisation (see for example \cite{ginsparg}) as
\beq
\mbox{det}' \bx=\tau_2\eta^2(\tau)\bar{\eta}^2(\tb),
\eeq
where $\eta(\tau)$ is the Dedekind $\eta$-function,
\beq
\eta(\tau)=e^{i\pi \tau/12}\prod_{n>1}(1-e^{2\pi i n \tau}).
\eeq
We now have the oscillator part of the partition function, given by
\beq\label{dZoscX}
Z_{osc}=\frac{R}{\sqrt{2\tau_2}|\eta|^2}.
\eeq
The contribution due to the $\xt$ functional integral is an identical factor with $R$ replaced by $1/R$, so the square root of the contribution of the doubled oscillators, which we expect to be, and is, the same as the constrained contribution, is given by
\beq\label{dZosc}
Z_{osc}=\frac{1}{\sqrt{2\tau_2}|\eta|^2}.
\eeq

To factorise the classical part of the partition function we worked in
terms of $P$ and $Q$ and used holomorphic factorisation, and we can
check that we get the same answer if we do that here. The substitution
(\ref{PQ}) introduces a Jacobian factor of $1/2$. Once the
substitution is made we do the path integral for the two Bosons, $P$ and $Q$,
just like the path integral for $X$ and $\xt$, except for a factor of
$1/2$ in the action and the more complex target space boundary
conditions that $P$ and $Q$ inherit as a result of their definition
(\ref{PQ}) in terms of $X$ and $\xt$. As the eigenfunctions of $\bx$,
(\ref{efns}), do not depend on these boundary conditions (unlike the
instanton pieces) the determinants for $P$ and $Q$ are the same as
those for  $X$ and $\xt$. However, the zero-mode integral does depend on the boundary conditions: although $P$ and $Q$ can take any value, the periodicity condition means we should only integrate over one fundamental region, we choose the one inherited from the fundamental region for $X$ and $\xt$, where $X$ and $\xt$ are allowed to range from 0 to 1. The volume of this region is given by an integral over the values $P$ can take, of the range of values $Q$ can take for that value of $P$, that is
\beq
\int_{P=0}^{R^{-1}}2PdP+\int_{P=R^{-1}}^{R}2R^{-1}dP+\int_{P=R}^{R+R^{-1}}2((R+R^{-1})-P)dP=2\, ,
\eeq
cancelling the factor from the Jacobian.

The normalisation factor (\ref{norm}) remains the same as the additional factor of $1/2$ on the action is cancelled by the Jacobian which should also be included in the normalisation integral (\ref{norm}) (or rather the root of the Jacobian as there is one Jacobian to be split between both the $P$ and the $Q$ normalisation integrals). The $P$ oscillator contribution is then 
\beq\label{Zoscp}
Z_{osc;P}=\frac{1}{\sqrt{2\tau_2}|\eta|^2}.
\eeq
The $Q$ contribution is identical, and again one can take the $\tau$ dependent holomorphic square root of the $P$ factor and the $\tb$ dependent anti-holomorphic square root of the $Q$ factor and multiply them together to again get (\ref{Zoscp}). Taking this together with (\ref{Zf}) we obtain that the partition function for a Boson of radius $R$ in the doubled formalism:
\beq\label{Z}
Z=\sum_{p_L,
p_R} \frac{1}{|\eta|^2}\exp\left[i\pi \tau \frac{p_L^2}{4}-i\pi\tb \frac{q_{R}^{2}}{4}\right]
\eeq
with
\beq
p_L=mR+\frac{n}{R}\, , \qquad q_R=mR-\frac{n}{R}\, .
\eeq
This is exactly what one gets for the same Boson using the undoubled formalism, as we will now calculate with our conventions.

\section{The Ordinary Boson}

In order to aid comparison with the result of the doubled formalism, 
we describe below the partition function of the ordinary Boson at one 
loop using appropriate conventions so as to compare results. We proceed 
in the same way as above. The action is
\beq
S=-\pi R^2 L\wedge *L\, ,
\eeq
with $L=dX+n\alpha+m\beta$, $m,n\in\Z$. We can then write the instanton sum part of the partition function as
\beq
Z_{inst}=\sum_{m,n} \exp \left[ -\pi R^2\left(n^2\frac{|\tau|^2}{\tau_2}-2mn\frac{\tau_1}{\tau_2}+\frac{m^2}{\tau_2}\right)\right]\, .
\eeq
Poisson resummation on $m$ gives
\bea\label{Bpf}
Z_{inst}&=&\sum_{n,w} \sqrt{\frac{\tau_2}{R^2}}\exp \left[ -\pi R^2\frac{n^2|\tau|^2}{\tau_2}-\frac{\tau_2\pi}{R^2}\left(w-\frac{in\tau_1R^2}{2\tau_2}\right)^2\right]\nonumber\\
&=&\sum_{n,w} \sqrt{\frac{\tau_2}{R^2}}\exp \left[ -\pi\tau_2\left(R^2n^2+\frac{w^2}{R^2}\right)+2\pi i n w \tau_1\right]\nonumber\\
&=&\sum_{n,w} \sqrt{\frac{\tau_2}{R^2}}\exp \left[ i\pi\tau \frac{p_L^2}{2}-i\pi\tb \frac{p_R^2}{2}\right]\, ,
\eea
where
\beq
p_L=Rn+\frac{w}{R}\, ,\qquad p_R=Rn-\frac{w}{R}\, .
\eeq
Performing Poisson resummation on $n$, rather than $m$, leads to a the same
result up to a modular transformation 
taking $\tau\rightarrow-1/\tau$. 

Evaluation of the oscillator part of the partition function proceeds
much as the previous section, leading up to (\ref{dZoscX}). The only
difference is that there is no factor of $1/\sqrt{2}$ due to the standard normalisation of the action, giving
\beq
Z_{osc}=\frac{R}{\sqrt{\tau_2}|\eta|^2},
\eeq
leading to the full partition function
\beq
Z=\sum_{m,n} \frac{1}{|\eta|^2}\exp \left[ i\pi\tau \frac{p_L^2}{2}-i\pi\tb \frac{p_R^2}{2}\right]\, ,
\eeq
where
\beq
p_L=Rn+\frac{w}{R}\, ,\qquad p_R=Rn-\frac{w}{R}\, .
\eeq
The partition function is now invariant for $R\rightarrow 1/R$, after
the relabelling of sums. In general there will be an $R$-dependent
factor outside the exponential which is absorbed into the dilaton
shift, but in the case of the torus there is no dilaton shift due to the vanishing Euler character. In our doubled calculation there was also no $R$-dependence in
the partition function, but both the instanton and oscillator pieces
were separately independent of $R$. For higher genus we expect
both pieces of the doubled partition function to remain $R$-independent, but for the ordinary Boson the instanton part will give
higher powers of $R$ whereas the $R$-dependence of the oscillator
part will remain the same (this contribution effectively comes from
the volume of the zero-mode but we have scaled into the target space
metric). This R-dependence will give the dilaton shift which is not
present in the doubled formalism, as here perturbation theory is in
terms of a differently defined, T-duality invariant, dilaton\cite{Hull:2006va}.

\section{Discussion}

We have shown the equivalence of the partition sum in Hull's
formulation with that of the ordinary string. Crucial to this was the inclusion
of the topological term introduced by Hull in \cite{Hull:2006va}. We have not
included all the subtleties of quantising chiral Bosons. In particular
we have not discussed the possibility,
crucial for nontrivial T-folds, of a non-global choice of
polarisation. This is left for future work.

\section*{Acknowledgements}
DSB is supported by EPSRC grant GR/R75373/02. There is also partial support 
from EC Marie Curie Research Training Network, MRTN-CT-2004-512194. 
DSB thanks DAMTP and in particular Clare Hall college Cambridge for 
continued support. We wish to thank Chris Hull for discussions on T-folds, 
Mans Henningson for discussions of holomorphic factorisation and Peter 
West on earlier formulations of duality symmetric theories.

\newpage
\section{Appendix: The Constraint}

In the formalism of Hull, toroidally compactified dimensions (or fibres of torus fibrations) are doubled, which elucidates how T-duality acts on the theory. This, of course, doubles the number of degrees of freedom: to ensure no additional physical degrees of freedom are introduced a constraint is required to maintain the original number. For one flat compactified dimension $X$ (with dual dimension $\xt$) this constraint is that $dX+d\xt$ be self-dual, and that $dX-d\xt$ be anti-self-dual. This can be re-expressed as $X+\xt$ is left-moving, or, once we have moved to a Euclidean world sheet, that $X+\xt$ is a holomorphic function of the complex world sheet co-ordinate $z=\sigma_1+\tau\sigma_0$, where $\tau$ is the complex structure of the toroidal worldsheet.

The general form of the constraint (in the absence of sources) is
\beq
\P^{I}=L^{IJ}\H_{JK}*\P^K.
\eeq
$\P^{I}$ are the doubled momenta, which can locally be written as $\P^I=d\X^{I}$, where $\X^I$ are the doubled torus, $I,J,\ldots$ run over the $2n$ co-ordinates of the doubled torus (note that we have not yet picked a polarisation, so there is no distinction over which half of the co-ordinates are fundamental, and which are dual). $\H_{IJ}$ is the positive metric on the doubled torus, and $L_{IJ}$ is an $O(n,n)$ invariant metric and $*$ is the world sheet Hodge dual.
We also need to introduce the vielbein, $\V$, such that
\beq
\H_{IJ}=\V^{t\ \, A}_{\ I}\delta_{AB}\V^B_{\ \, J}\, ,
\eeq
where $A,B,\ldots$ are $O(n)\times O(n)$ indices which split into the two factors as $A=(a,a')$. In the $O(n)\times O(n)$ basis $L^{AB}$ can be written
\beq
L^{AB}=\left(\begin{array}{cc}\openone^{ab} & 0\\ 0&-\openone^{a'b'}\end{array}\right)
\eeq
so in terms of $\P^A=\V^A_{\ \,I}\P^I$ the constraint becomes
\bea\label{Pconst}
\P^a&=&*\P^a\non\\
\P^{a'}&=&-*\P^{a'}.
\eea
As long as we are dealing with a trivial bundle this implies that, in terms of the light cone co-ordinates on the world sheet, $\partial_-X^a=0$ and $\partial_+X^{a'}=0$. Hence the relationship with the chiral Boson. After Wick rotation these conditions become $\partial_{\zb} X^a=0$ and $\partial_z X^{a'}=0$, allowing us to use a form of holomorphic factorisation.

We now turn to our example where we have one toroidally compactified dimension, $X$, of constant radius $R$. We introduce a dual co-ordinate, $\xt$, of radius $R^{-1}$.  Choosing $X$ as the original co-ordinate means we are working with a specific polarisation, that is we have chosen a basis where the co-ordinates separate into the fundamental and the dual representations of $GL(n,{\mathbb R})$ ($n=1$ here), labelled by $\ti=(\, ^i,\, _i)$. The projectors 
\beq
\Phi^{\ti}_{\ I}=\left(\begin{array}{c}\Pi^{i}_{\ I}\\ \Pi_{i \,  I}\end{array}\right)
\eeq
take us to this basis and
\bea
X&=&X^i=\Pi^{i}_{\ I}\X^I, \non\\
\xt&=&X_{i}=\Pi_{i\,  I}\X^I. 
\eea
Also in this basis we have
\beq
\H_{\ti\tj}=\left(\begin{array}{cc}R^2 & 0\\0&R^{-2}\end{array}\right)
\eeq
and
\beq
L_{\ti\tj}=\left(\begin{array}{cc}0&1\\1&0\end{array}\right)\, .
\eeq
We introduce $K^A_{\ \ti}$ via $ K=\V\Phi^t$ which will allow us to relate things in the $O(n,n)$ basis (where the constraint is simplest) to the $GL(n,{\mathbb R})$ basis (where our original Boson can be seen). As we know $L$ and $\H$ in both bases we can determine that $K$ is given by\footnote{We have made some sign choices which do not affect the constraint.}
\beq
K^A_{\ \ti}=\frac{1}{\sqrt{2}}\left(\begin{array}{cc}R & R^{-1}\\ R&-R^{-1}\end{array}\right)
\eeq
This means that $\P^a$ appearing in the constraint (\ref{Pconst}) is related to $\P^i=dX$ and $\P_i=d\xt$ via $\P^a=K^a_{\ \ti}\P^{\ti}$, giving
\beq\label{Pa}
\P^a=\frac{1}{\sqrt{2}}\left(RdX+R^{-1}d\xt\right)\, ,\P^{a'}=\frac{1}{\sqrt{2}}\left(RdX-R^{-1}d\xt\right)\, .
\eeq
We will work in terms of $P$ and $Q$ where $\sqrt{2}\P^a=dP$ and $\sqrt{2}\P^{a'}=dQ$. As a consequence of (\ref{Pa}) they obey
\bea
\partial_{\zb} P&=&0\, ,\\
\partial_{z} Q &=&0\, .
\eea


\newpage

\end{document}